\documentclass[twocolumn,showkeys,preprintnumbers,amsmath,amssymb,superscriptaddress]{revtex4}

\usepackage{graphicx}
\usepackage{dcolumn}
\usepackage{bm}
\usepackage{tabularx}
\usepackage{color}

\begin{document}

\title{Invariant Wide Bandgaps in Honeycomb Monolayer and Single-Walled Nanotubes of IIB-VI Semiconductors}

\author{Xiaoxuan Ma}
\affiliation{College of Physics, Optoelectronics and Energy, Soochow University, Suzhou, Jiangsu 215006, People's Republic of China}

\author{Jun Hu}
\email{jhu@suda.edu.cn}
\affiliation{College of Physics, Optoelectronics and Energy, Soochow University, Suzhou, Jiangsu 215006, People's Republic of China}

\author{Bicai Pan }
\email{bcpan@ustc.edu.cn}
\affiliation{Department of Physics and Hefei National Laboratory for Physical Sciences at Microscale, University of Science and Technology of China, Hefei, Anhui 230026, People's Republic of China}


\begin{abstract}
Search for low-dimensional materials with unique electronic properties is important for the development of electronic devices in nano scale. Through systematic first-principles calculations, we found that the band gaps of the two-dimensional honeycomb monolayers and one-dimensional single-walled nanotubes of IIB-VI semiconductors (ZnO, CdO, ZnS and CdS) are nearly chirality-independent and weakly diameter-dependent. Based on analysis of the electronic structures, it was found that the conduction band minimum is contributed by the spherically symmetric $s$ orbitals of cations and the valence band maximum is dominated by the in-plane $(d_{xy}-p_y)$ and $(d_{x^2-y^2}-p_x)$ hybridizations. These electronic states are robust against radius curvature, resulting in the invariant feature of the band gaps for the structures changing from honeycomb monolayer to single-walled nanotubes. The band gaps of these materials range from 2.3 eV to 4.7 eV, which is of potential applications in electronic devices and optoelectronic devices. Our studies show that searching for and designing specific electronic structures can facilitate the process of exploring novel nanomaterials for future applications.
\end{abstract}

\keywords{Wide Bandgap, Honeycomb Monolayer, Single-Walled Nanotube, Semiconductor}


\maketitle

\section {Introduction}

Advances in modern technologies accelerate miniaturization of electronic devices, which drives the endeavors for exploring exotic materials in nano scale. In the last two decades, a lot of low-dimensional materials such as clusters \cite{Rohlfing}, nanotubes \cite{Iijima}, and atomically thin films \cite{graphene,MoS2} have been discovered. These low-dimensional materials exhibit intriguing and abundant physical properties, ranging from metallic conductor to semiconductor, which guarantee them promising wide applications in electronic and optoelectronic devices in nano scale \cite{XiaYN,WangZL,LieberCM,YangPD,CaoJL}.

As the firstly discovered one-dimensional material \cite{Iijima}, carbon single-walled nanotubes (SWNTs) were investigated extensively for the possible applications in new generation of electronic devices. However, the electronic properties of the carbon SWNTs depend on their chirality and diameter. For example, the ($n,\ m$) carbon SWNTs are metallic when $n-m=3l$ ($l$ is an integer), while the others are semiconducting \cite{Hamada,Lieber,Charlier}. This largely hinders the application of carbon SWNTs, because it is difficult to repeatedly fabricate carbon SWNTs with exactly the same chirality and diameter, so that all the products (i.e. carbon SWNTs) have the same electronic property. Therefore, invariably semiconducting SWNTs such as boron nitride (BN) SWNTs are more favorable for the applications in electronic devices \cite{Louie1,Rubio1,XiangHJ,Baumeier}. Interestingly, the band gaps of armchair ($n,\ n$) BN SWNTs are almost constant and independent of the diameters, while those of zigzag ($n,\ 0$) BN SWNTs decrease as the diameters decrease \cite{Louie1,XiangHJ}. These electronic features are mainly attributed to the $sp^2\sigma+pp\pi$ hybridizations of the valence orbitals \cite{Charlier,Baumeier}. The $sp^2\sigma$ hybridization mainly contributes the strong in-plane chemical bonds, while the $pp\pi$ hybridization contributes the electronic states near the Fermi energy. Therefore, the electronic properties of the carbon and BN SWNTs are mainly dominated by the  $pp\pi$ hybridization. It is known that the $pp\pi$ hybridization occurs between the $p_z$ orbitals. Since the spatial distribution of the $p_z$ orbital is perpendicular to the cylinder surface of the carbon and BN SWNTs, the $pp\pi$ hybridization is sensitively affected by the chirality and diameter of the carbon and BN SWNTs. Consequently, the electronic properties of the carbon and BN SWNTs are chirality- and diameter-dependent.

Recently, the SWNTs of wide-bandgap semiconductor ZnO were studied extensively \cite{TuZC,Mintmire,ZhaoJJ,ZhangRQ,ShenX,Marana}. It was found that the band gaps of ZnO SWNTs are almost insensitive to the chirality and diameter, being significantly different from the carbon and BN SWNTs. However, the microscopic origin of this feature in ZnO SWNTs is still unclear. Furthermore, no SWNTs of wide-bandgap semiconductor have been synthesized in experiment so far. Nevertheless, recent first-principles calculations predicted that the ultrathin (0001) surface of wurtzite IIB-VI and III-V semiconductors prefers to adopt honeycomb lattice due to the electrostatic interaction between cation and anion layers \cite{Freeman}. This prediction was confirmed by experiments for the cases of ZnO \cite{Tusche} and GaN \cite{GaN}. These achievements are the precursors for fabricating SWNTs of wurtzite IIB-VI and III-V semiconductors, because the SWNTs are the transformation of honeycomb monolayer (HM) by rolling up the later \cite{CastroNeto}. Therefore, the HMs and SWNTs of wurtzite IIB-VI and III-V semiconductors may be produced in experiment under proper condition and applied in electronic devices in the future. Clearly, revealing the electronic features of the HMs and SWNTs of these semiconductors is useful and important for the design of these materials in electronic devices.

In this paper, taking ZnO, CdO, ZnS and CdS as the prototypes of IIB-VI semiconductors, we studied the stabilities and electronic properties of the HMs and SWNTs, by using first-principles calculations. We found that both the HMs and SWNTs are structurally stable, and their band gaps are insensitive of their chirality and diameter. The analysis of the electronic structures revealed that the $d$ orbital of Zn and Cd plays crucial roles in the electronic characteristic. 

\section {Computational details}
All calculations were performed by using the first-principles pseudopotential method based on density functional theory (DFT) within local density approximation (LDA) as implemented in SIESTA package \cite{Siesta}. The pseudopotentials were constructed by the Troullier-Martins scheme \cite{TM2}. The Ceperley-Alder exchange-correlation functional \cite{CA} as parameterized by Perdew and Zunger  \cite{PZ} was employed. In our calculations, the double-$\zeta$ plus polarization basis sets were chosen for all atoms. For the HMs and SWNTs, the $20\times20\times1$ and $1\times1\times20$ $k$-grid meshes within the Monkhorst-Pack scheme \cite{Monkhorst} in the Brillouin zone were considered, respectively. The atomic structures were fully relaxed using the conjugated gradient method until the Hellman-Feynman force on each atom is smaller than 0.01 eV/{\AA}. The generalized gradient approximation (GGA) with PBE functional \cite{PBE} within SIESTA package and B3LYP hybrid functional \cite{crystal} within CRYSTAL03 package were also employed to explore the effect of different functionals on the electronic properties. In the B3LYP calculations, the Stuttgart-Dresden effective core pseudopotentials (ECP) \cite{ECP} were used.

\section {Results and discussions}
\subsection{Wurtzite bulk versus honeycomb monolayer}

\begin{figure}
\centering
\includegraphics[width=8.5cm]{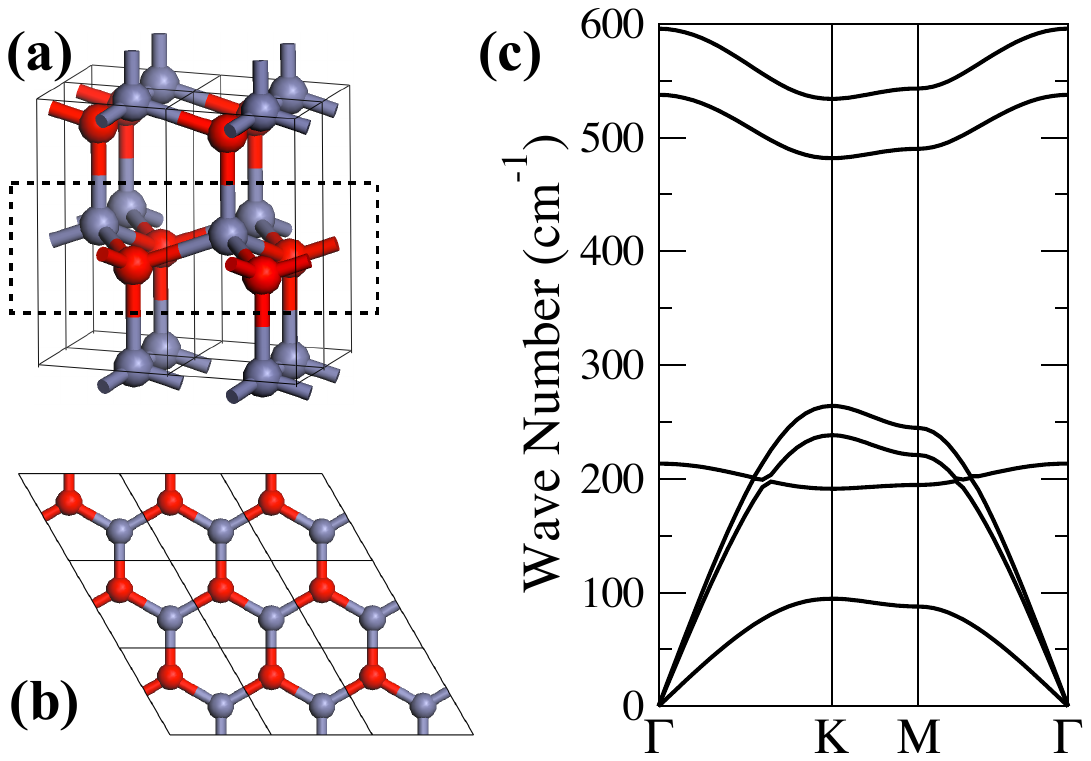}
\caption{(Color online) (a) Wurtzite and (b) honeycomb monolayer of ZnO. The red and gray spheres stand for O and Zn atoms, respectively. (c) Phonon dispersion curves of honeycomb monolayer ZnO.
}\label{fig-vibrate}
\end{figure}

Most IIB-VI and III-V compounds crystallize wurtzite (WZ) structure as shown in Fig.~\ref{fig-vibrate}(a). The cations and anions stack layer by layer alternatively along the (0001) direction. If we cut a two-layer slab composed of one cation layer and one anion layer as indicated by the dashed rectangle in Fig.~\ref{fig-vibrate}(a), the Coulomb interaction between them transforms the slab into planar honeycomb structure \cite{Freeman}, as shown in Fig~\ref{fig-vibrate}(b).  Therefore, we chose ZnO, CdO, ZnS and CdS as the prototypes of wurtzite IIB-VI semiconductors to investigate the structural and electronic properties of their HMs. We firstly optimized the atomic structures of ZnO, CdO, ZnS and CdS in WZ and HM phases. As listed in Table I, the bond length between cation and anion in HM phase is shorter by $3\%\sim5\%$ than that in the corresponding WZ phase, due to the reduction of dimensions of the materials. In addition, the binding energies of all considered cases are negative, indicating that the interaction between cations and anions are energetically exothermal. For each compound, the binding energy of the HM phase is higher than that of the WZ phase, which is reasonable because the WZ is the ground state phase. To further investigate the stability of the HM phase of these compounds, we calculated the phonon dispersions of ZnO HM by using the ``frozen phonon" approach \cite{vibrate}. As shown in Fig.~\ref{fig-vibrate}(c), the optical and acoustical branches are well separated and all branches have positive frequency. Therefore, the HM phase is a metastable phase of ZnO, in agreement with the experimental observation \cite{Tusche}. This conclusion can be extended to other compounds considered in this work.

\begin{table}
 \centering
 \caption{Bond length (in {\AA}) between cations and anions, and binding energy ($E_b$, in eV per formula unit), and band gap ($E_g$, in eV) of ZnO, CdO, ZnS and CdS in WZ and HM. $\Delta$ is the contraction of bond lengths in HM compared to that in WZ. The values in parentheses are experimental band gaps.
}
 \tabcolsep0.05in             
 \begin{tabular}{cccccccccc}
   \hline
   \hline
   & \multicolumn{3}{c}{Bond length} & & \multicolumn{2}{c}{$E_b$} & & \multicolumn{2}{c}{$E_g$}\\
   \cline{2-4} \cline{6-7} \cline{8-10}
     & WZ & HM & $\Delta$(\%) & & WZ & HM  & & WZ & HM \\
   \hline
   ZnO & 1.99 & 1.90 & 4.62 & & -9.19 & -8.26 & & 0.94 (3.44) & 2.03 \\
   CdO & 2.18 & 2.10 & 3.44 & & -7.81 & -6.82 & & 0.00 (0.84) & 0.77 \\
   ZnS & 2.34 & 2.23 & 4.95 & & -7.69 & -6.81 & & 2.19 (3.91) & 2.79 \\
   CdS & 2.53 & 2.42 & 4.46 & & -6.82 & -5.93 & & 0.96 (2.48) & 1.64 \\
   \hline
   \hline
 \end{tabular}
\end{table}

Then we calculated the band gaps of all cases as listed in Table I. Clearly, the band gap ($E_g$) of HM phase is significantly larger than that of WZ phase for each compound. For example, the $E_g$ of ZnO HM increases by about 1.1 eV, from 0.94 eV in WZ ZnO to 2.03 eV. Note that the $E_g$ of WZ ZnO from our LDA calculation is much smaller than the experimental value (3.4 eV) \cite{FengXB,ZnO-gap}, because the DFT calculations usually underestimate the band gaps of semiconductors. It is even worse for WZ CdO of which the calculated $E_g$ is less than 1 meV, in agreement with previous theoretical reports \cite{Roberto,ZhuYZ}, but obviously wrong since CdO is a semiconductor with indirect $E_g$ of 0.84 eV and direct $E_g$ of 2.28 eV \cite{CdO-gap}.  This problem may be relieved by using hybrid functional such as B3LYP \cite{HuJun}, which will be discussed in more details later. Nonetheless, the significant widening of the band gaps is still qualitatively reasonable, since it originates from the change of geometric symmetry and quantum confinement effect.

\begin{figure}
\centering
\includegraphics[width=8.5cm]{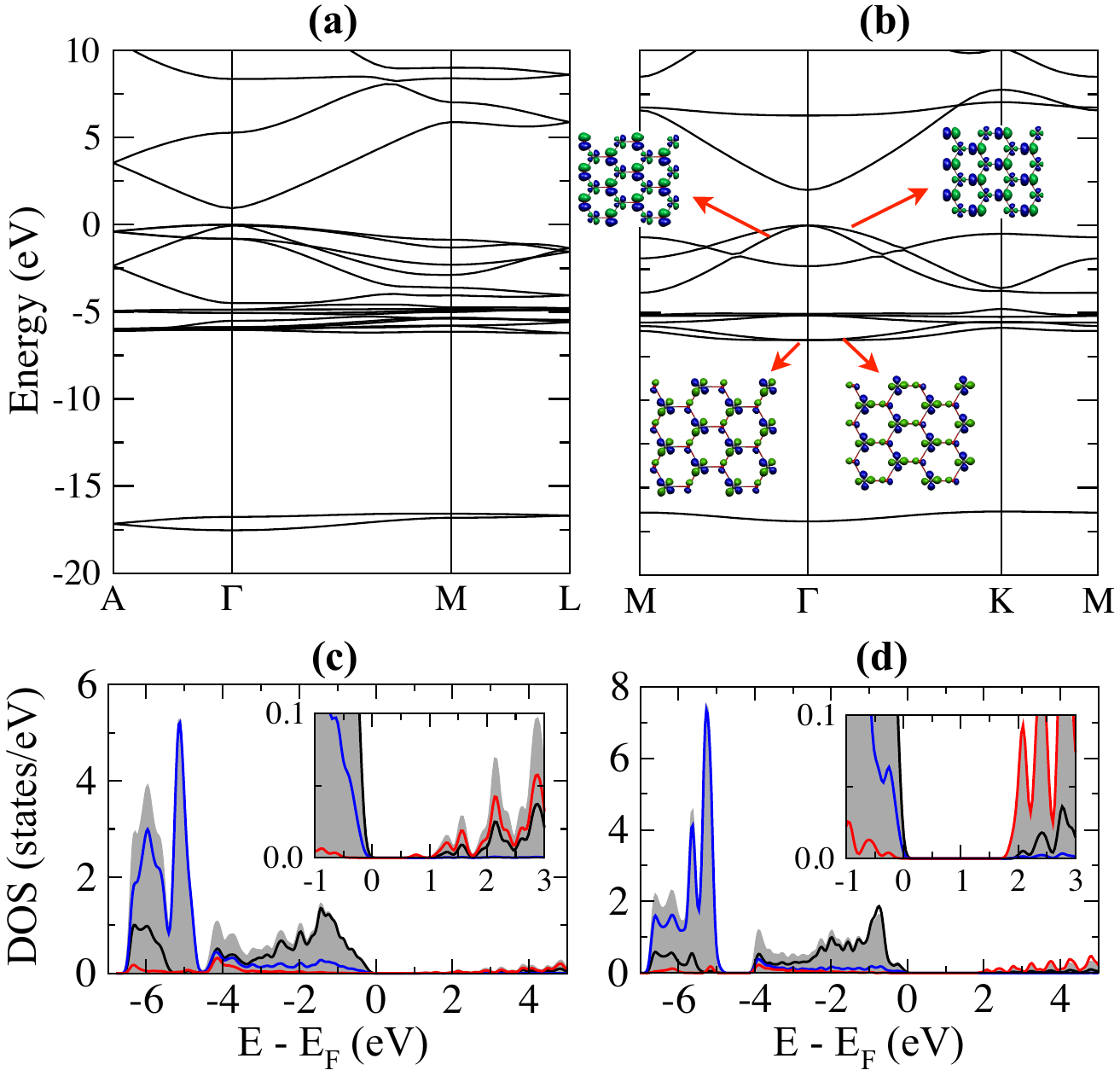}
\caption{(Color online) Band structures and density of states (DOS) of ZnO in WZ phase [(a) and (c)] and HM phase [(b) and (d)]. The valence band maximum is set to zero point of energy. The insets in (b) are the spatial distribution of the radial wavefunctions of the energy levels indicated by the red arrows at $\Gamma$ point. These energy levels correspond to the bonding and anti-bonding states of $(d_{xy}-p_y)$ (left) and $(d_{x^2-y^2}-p_x)$ (right) hybridizations, respectively. The cutoff of the isosurfaces is 0.1 electrons/{\AA}$^3$.  The red and gray spheres stand for O and Zn atoms, respectively. The green and blue isosurfaces represent positive and negative wave functions. The gray areas in (c) and (d) are the total DOS. The black, red and blue curves are the projected density of states of O$-2p$, Zn$-4s$ and Zn$-3d$ orbitals, respectively.
}\label{fig-zno-band}
\end{figure}

To reveal the electronic feature of these HMs, we chose ZnO as prototype and plotted the band structures and density of states (DOS) in WZ and HM phases in Fig. \ref{fig-zno-band}. From the band structures in Fig. \ref{fig-zno-band}(a) and Fig. \ref{fig-zno-band}(b), it can be seen that both WZ and HM ZnO are direct-band gap semiconductors, with the valence band maximum (VBM) and conduction band minimum (CBM) at $\Gamma$ point. However, the band dispersions are significantly different. Furthermore, a gap from $-5$ eV to $-4$ eV appear in the band structure of HM phase. From the density of states (DOS) in Fig. \ref{fig-zno-band}(c) and \ref{fig-zno-band}(d), the states from $-6.5$ eV to 5 eV are mainly contributed from Zn$-3d$ orbitals, the states ranging $-4\sim0$ eV are from O$-2p$ orbitals, and the states near the CBM are from Zn$-4s$ orbital.

In the WZ structure, any atom locates at the center of the tetrahedron composed of four neighboring atoms of the other type of element. Therefore, the $s$ and $p$ orbitals of anions adopt the $sp^3$ hybridization and then hybridize with all components of the $d$ orbitals of cations. On the contrary, in the HM phase, the $s$ and $p$ orbitals of anions adopt the $sp^2$ hybridization, and the $d$ orbitals of cations split into three group: $d_{z^2}$, $d_{xz/yz}$ and $d_{xy/x^2-y^2}$. Consequently, hybridizations in HM phase should be significantly different from those in WZ phase. To reveal the role of hybridizations in the electronic properties, we calculated the radial wavefunctions of the energy levels at $\Gamma$ point. The VBM of ZnO HM is doubly degenerate and mainly originates from the anti-bonding states of the in-plane $(d_{xy}-p_y)$ and $(d_{x^2-y^2}-p_x)$ hybridizations, as shown in the insets in Fig. \ref{fig-zno-band}(b). The corresponding bonding states are 6.6 eV bellow, manifesting the strong interactions between O$-2p_{x/y}$ and Zn$-3d_{xy/x^2-y^2}$ orbitals. In addition, the energy level at $-2.3$ eV ($\Gamma$ point) is mainly contributed from O$-2p_z$ orbital, mixed with a little part of Zn$-4p_z$ orbital. This state represents weak $pp\pi$ hybridization between O$-2p_z$ and Zn$-4p_z$ orbitals. The narrow bands near $-5.0$ eV are from $d_{z^2}$ and $d_{xz/yz}$ of Zn atom, implying that the $d_{z^2}$ and $d_{xz/yz}$ orbitals maintain atomic orbital feature and do not hybridize with O$-2p$ orbitals. Therefore, the in-plane $(d_{xy}-p_y)$ and $(d_{x^2-y^2}-p_x)$ hybridizations dominate both the chemical bonds and electronic states around the Fermi energy, which is different from the electronic nature of C and BN monolayers. The electronic structures of other compounds considered in this work have similar feature as ZnO, but the relative positions of the energy levels are different due to the different atomic sizes and bond lengths (see Table I).

\subsection{Single-walled nanotubes}

\begin{figure}
\centering
\includegraphics[width=6.5cm]{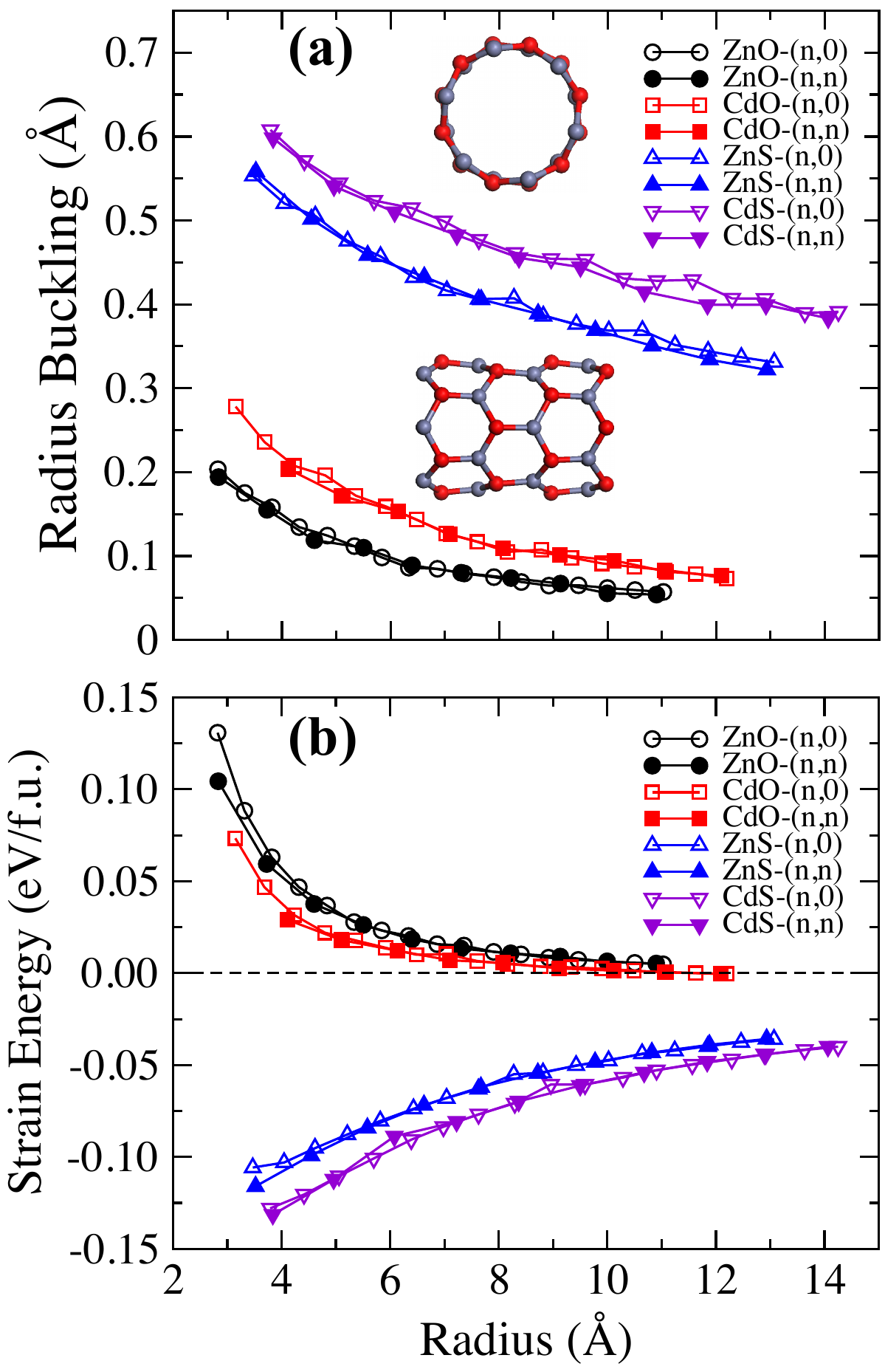}
\caption{(Color online) (a) Radius buckling (${R_{anion}-R_{cation}}$) and (b) strain energy of all considered SWNTs. Insets in (a) are top and side views of (6,0) ZnO SWNT. The red and gray spheres stand for O and Zn atoms, respectively. 
}\label{fig-structure}
\end{figure}

The atomic structures of the SWNTs of ZnO, ZnS, CdO and CdS are similar to the BN SWNTs, with cations (Zn and Cd) and anions (O and S) replacing the B and N atoms, respectively. For all compounds, we  considered zigzag SWNTs from (5, 0) to (21, 0) and armchair SWNTs from (3, 3) to (12, 12), with the radius varying from 2.9 to 14.5 {\AA}. Both the atomic positions and the axial lattice constants are optimized. In these SWNTs, the walls are buckled, with the outer and inner cylinders composed of anions and cations, respectively. From Fig.~\ref{fig-structure}(a), it can be seen that the amplitudes of the buckling are dependent on the radii of the SWNTs but independent of the chirality. For each compound, the smaller the radius is, the larger the buckling is. Interestingly, the overall radius buckling of the oxide SWNTs is much smaller than that of the sulfide SWNTs. The buckling can be as large as 0.6 {\AA} for small CdS SWNTs [e.g. (5,0) and (3,3)], and still be $\sim$ 0.4 {\AA} for those with radii of $\sim$ 14 {\AA}. On the other hand, the buckling is only $\sim$ 0.2 {\AA} for (5,0) and (3,3) ZnO SWNTs whose radii are smaller than 3 {\AA}, and decreases to $\sim$ 0.05 {\AA} when the radius reaches 11 {\AA}. This phenomenon originates mainly from the different ionicity of oxide and sulfide compounds. In fact, the ionicity of oxide compound (ZnO and CdO) is much stronger than that of sulfide compounds (ZnS and CdS). Therefore, the Coulomb interaction in oxide compounds is stronger than that in sulfide compounds, which results in smaller radius buckling between cation and anion cylinders.

Usually, rolling up a HM to form a SWNT requires extra energy which is defined as strain energy ($E_s$) \cite{ZhaoJJ},
\begin{equation}
    E_s=E_{SWNT}-E_{HM},
    \label{E_s}
\end{equation}
where $E_{SWNT}$ and $E_{HM}$ are the total energies per formula unit (f.u.) of a SWNT and a HM, respectively. Usually, the probability of the formation of SWNTs depends on their strain energies: smaller strain energy corresponds to larger probability. As shown in Fig.~\ref{fig-structure}(b), the amplitudes of the strain energies are almost independent of the chirality for all the SWNTs. Consequently, the probabilities of the formations of zigzag and armchair SWNTs of the same compound are nearly the same, if their diameters are the same. For the oxide SWNTs, the strain energies are positive, which implies that the oxide SWNTs are less stable than the corresponding HMs. In addition, the $E_s$ decreases as the radius increases, and approaches to zero for large SWNTs. On the contrary, the strain energies of sulfide SWNTs are negative, which indicates that the process of rolling up a sulfide HM is energetically exothermic. Therefore, a sulfide HM may not exist, it rolls up to form a SWNT spontaneously. As a consequence, it may be easier to fabricate sulfide SWNTs than oxide SWNTs.

\begin{figure}
\centering
\includegraphics[width=8.5cm]{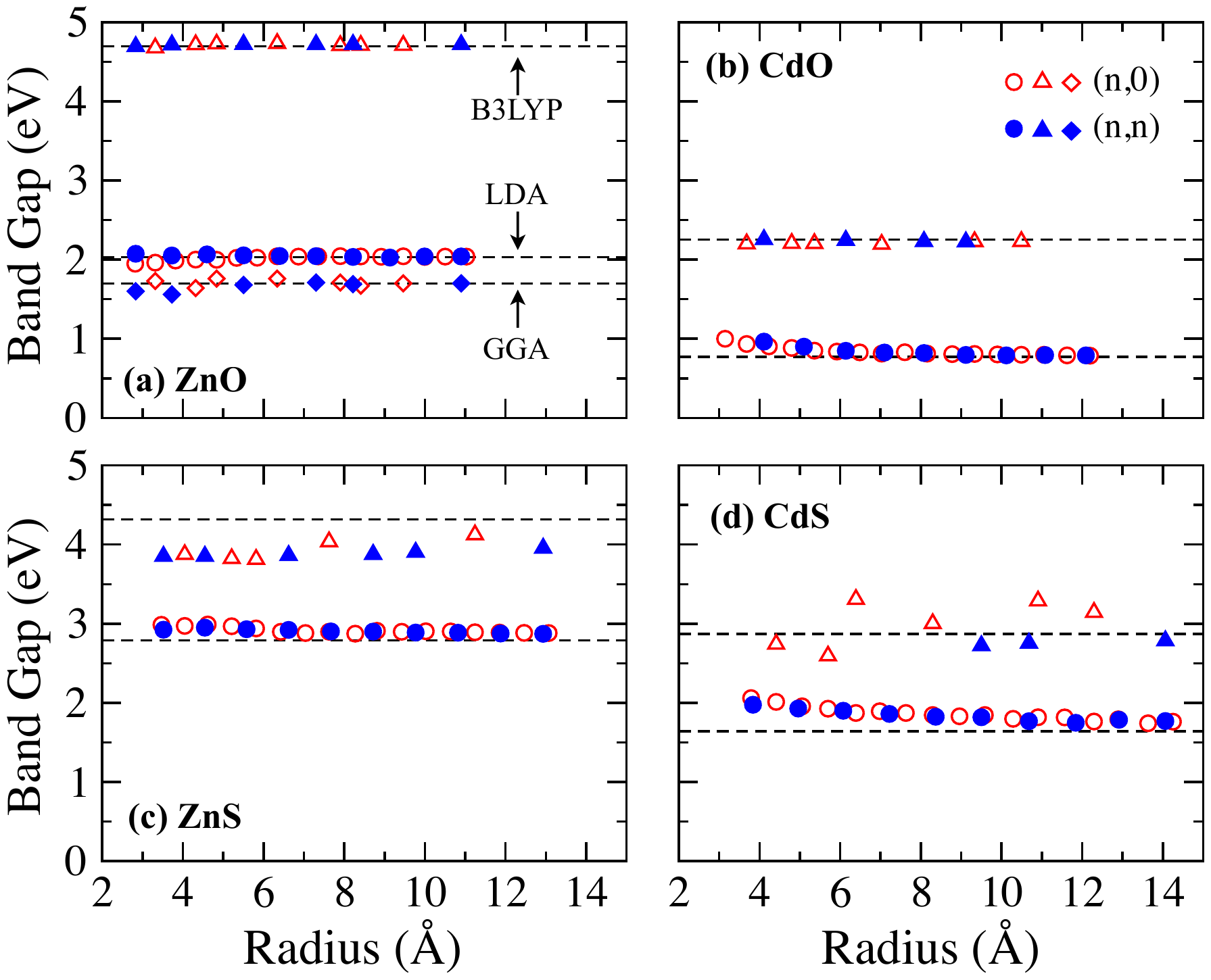}
\caption{Band gaps of zigzag (open symbols) and armchair (filled symbols) SWNTs.The circles, diamonds and triangles are from LDA, GGA and B3LYP calculations. The horizontal dashed lines represent the band gaps of the HMs with different functionals.
}\label{fig-band-gap}
\end{figure}

Then we calculated the band structures of all SWNTs and plotted the band gaps in Fig.~\ref{fig-band-gap}. Obviously, it can be seen that all SWNTs are  semiconducting. The band gaps of CdO, ZnS and CdS SWNTs are larger than that of the corresponding HM and decrease as the radii increase. The maximum deviations of the band gaps from those of the HMs are 0.23, 0.20 and 0.42 eV, respectively for CdO, ZnS and CdS SWNTs. Nonetheless, these deviations are much smaller than other kinds of SWNTs such as BN and SiC SWNTs in literature (about $1-3$ eV) \cite{XiangHJ,Baumeier,ZhaoMW,GuoGY}. Furthermore, the band gaps of these SWNTs are independent of the chirality. On the contrary, the band gaps of ZnO SWNTs with radii smaller than 6 {\AA} depend on the chirality: the band gaps of the zigzag SWNTs are smaller than that of the HM ZnO, whereas the band gaps of the armchair SWNTs are larger than that of the HM ZnO. However, the deviations of the band gaps from that of the HM ZnO is quit small, with the maximum difference of only 0.08 eV and the band gaps of the ZnO SWNTs with radii larger than 6 {\AA} are almost the same as that of HM ZnO. We should point out that the invariant character of the band gaps of these materials implies that it is not necessary to exactly control the chirality and diameter of the SWNTs to obtain one-dimensional semiconductors with the same band gap. Therefore, opposite to the carbon SWNTs whose chirality- and diameter-dependent electronic property hinders  their applications in the electronic devices, the oxide and sulfide SWNTs considered in this work are very promising for future applications in electronic devices, because their particular electronic property affords large flexibility for the process of producing these SWNTs as building blocks of electronic devices.

It is known that DFT calculations with conventional functionals such as LDA and GGA usually underestimate the band gaps of semiconductors. Fortunately, hybrid functional B3LYP can significantly relieve the band-gap problem. For example, the $E_g$ of WZ ZnO from B3LYP calculation was predicted as 3.41 eV \cite{HuJun} which reproduces the experimental measurement \cite{FengXB,ZnO-gap}. Therefore, we carried out calculations with B3LYP functional to obtain more accurate band gaps and further check whether the band gaps are still invariable. Interestingly, it can be seen from Fig. \ref{fig-band-gap} that the band gaps of oxide SWNTs are almost the same as the $E_g$ of the corresponding HM, around 4.7 and 2.3 eV for ZnO SWNTs and CdO SWNTs, respectively. The band gaps of ZnS are also steady but smaller by $0.2\sim0.5$ eV than the $E_g$ of HM ZnS (4.3 eV). For the CdS SWNTs, the band gaps vary within $\pm0.4$ eV with respective to the $E_g$ of HM CdS (2.9 eV). The larger deviations of the band gaps of the sulfide SWNTs relative to the oxide SWNTs originate from the larger radius buckling in the sulfide SWNTs. Nevertheless, the band gaps of the sulfide SWNTs are still quite close to those of the corresponding HMs. Furthermore, we used GGA functional to calculate the band gaps of the HM and SWNTs of ZnO and found that the band gaps are steadily around 1.7 eV as shown in Fig. \ref{fig-band-gap}(a). Therefore, the invariant feature of the band gaps of the oxide and sulfur SWNTs is predicted by LDA, GGA and B3LYP calculations. 

\begin{figure}
\centering
\includegraphics[width=8.5cm]{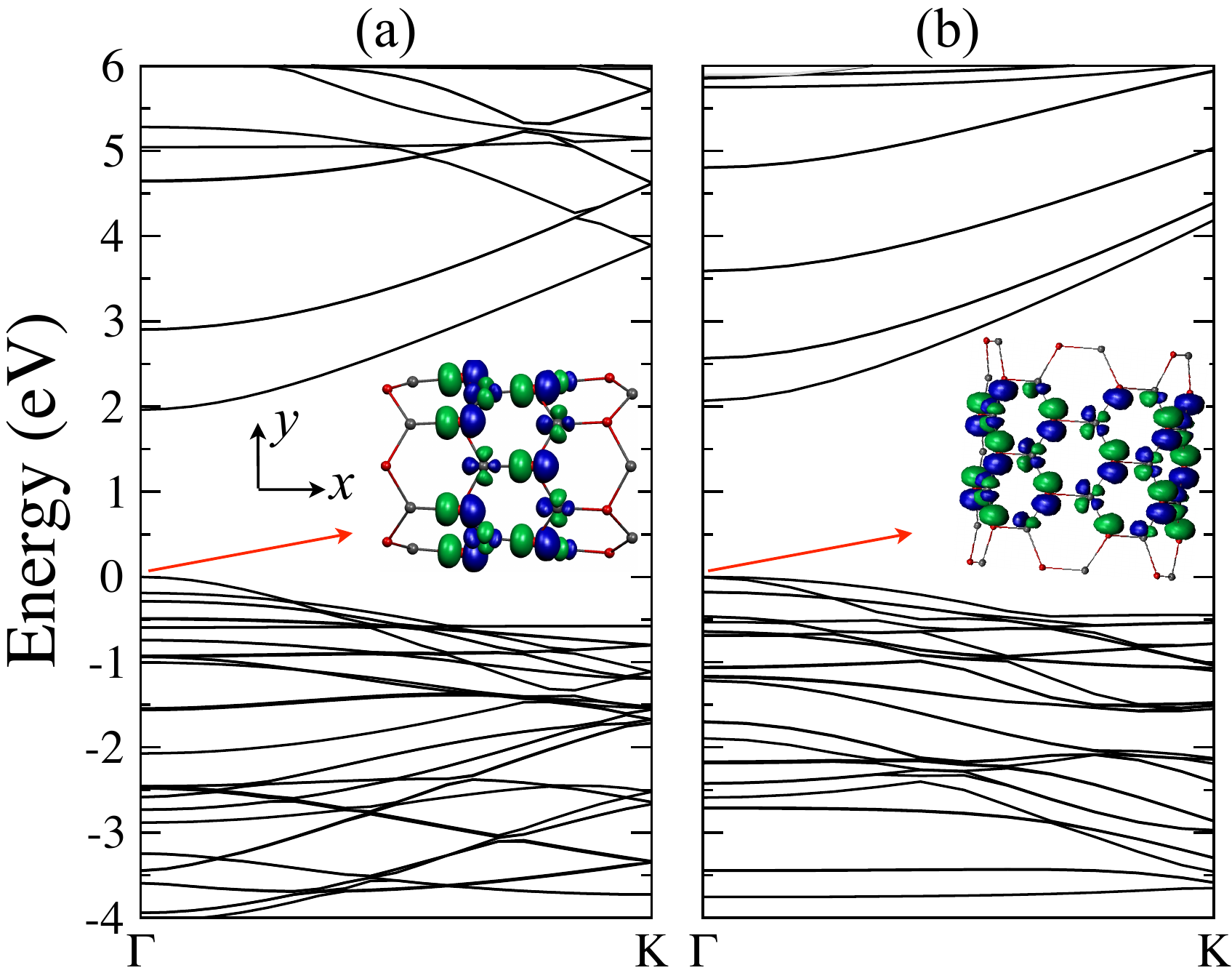}
\caption{(Color online) Band structures of (a) (6, 0) and (b) (5, 5) ZnO SWNTs. The VBM is set to zero point of energy. The insets show the spatial distribution of the radial wavefunctions of the VBM. The cutoff of the isosurfaces is 0.1 electrons/{\AA}$^3$. The red and gray balls stand for O and Zn atoms respectively.
}\label{fig-tube-band}
\end{figure}

We emphasize that our findings mentioned above are significantly different from the conventional SWNTs such as C, BN and SiC SWNTs where the $sp^2\sigma$ and $pp\pi$ hybridizations contribute the chemical bonds and the electronic states near the Fermi energy, respectively. These special hybridizations lead to the strongly chirality- and diameter-dependent electronic properties. For the IIB-VI compound SWNTs, however, the electronic properties are almost independent of the chirality and weekly dependent on the diameter. To illustrate the underlying mechanism, we investigated the electronic properties of these SWNTs. Take ZnO SWNTs as examples, we plotted the band structures of (6, 0) and (5, 5) ZnO SWNTs in Fig.~\ref{fig-tube-band}. Clearly, these SWNTs are direct-band gap semiconductors, with their gaps of 1.96 eV and 2.07 eV, respectively, very close to that of ZnO HM (2.03 eV). In addition, the CBM and VBM of both ZnO SWNTs locate at $\Gamma$ point. We found that the CBM is contributed mainly from the Zn$-4s$ orbital, and the VBM characterizes anti-bonding states of the $(d_{x^2-y^2-p_x})$ and $(d_{xy}-p_y)$ hybridizations [Fig.~\ref{fig-tube-band}]. Obviously, these states maintain the main electronic nature of HM ZnO [Fig. \ref{fig-zno-band}(b)], even though the ZnO HM is rolled up and radius curvature is induced. Both the spherically symmetric Zn$-4s$ orbital and the in-plane $(d_{x^2-y^2-p_x})^*$ and $(d_{xy}-p_y)^*$ hybridizations are robust against the radius curvature, which results in the constant band gaps of ZnO SWNTs. This mechanism applies to all the cases considered in this work.

\section {Summary}
In summary, we studied the stabilities and electronic properties of the HMs and SWNTs of IIB-VI semiconductors ZnO, CdO, ZnS and CdS, through systematic first-principles calculations. The sulfide SWNTs are easier to be fabricated than the oxide SWNTs, because the strain energies of the former are lower. Interestingly, the band gaps of the HMs and SWNTs of all the compounds are nearly chirality-independent and weakly diameter-dependent. This feature is contributed from the special electronic character of these materials. The CBM characterizes the $s$ orbitals of cations which are spherically symmetric, and the VBM originates from the in-plane $(d_{xy}-p_y)$ and $(d_{x^2-y^2}-p_x)$ hybridizations. The band gaps of these materials range from 2.3 eV to 4.7 eV, which make them suitable to be the building-block semiconductors in electronic devices and optoelectronic devices in nano scale.

\section {Acknowledgments}

This work is supported by the National Natural Science Foundation of China (11574223), the Natural Science Foundation of Jiangsu Province (BK20150303) and the Jiangsu Specially-Appointed Professor Program of Jiangsu Province. We also acknowledge the National Supercomputing Center in Shenzhen for providing the computing resources.


\end{document}